\documentstyle[12pt]{article}
\begin{document}
\noindent
\begin{center}
Bosonic stimulation and the irreproducibility of condensate fragmentation\\~\\
N. Kumar\\
Raman Research Institute, Bangalore 560080, India\\~\\
\end{center}
\begin{abstract}
It is pointed out that the quantum statistical phenomenon of Bosonic stimulation, inherent to Bose statistics and the associated Bose-Einstein correlation, can be effectively mapped on to the statistical problem of the Polya urn scheme. Thus, we predict an irreproducibility for the limiting non-degenerate values of the relative populations of two, or more equivalent but separated condensates resulting from the fragmentation of a given source condensate. Experiments are proposed that should verify this prediction $-$ using the dilute gas Bose-Einstein condensates, or a pulsed degenerate multi-mode laser, where one would look for the run-to-run (or the pulse-to-pulse) fluctuations of the relative populations.
\end{abstract}
It is a deep, but rather well known result that the indistinguishability of identical Bosons, and the associated quantum (Bose) statistics, implies a certain non-classical Bose-Einstein correlation.$^{1,2}$  Thus, the indistinguishable Bose particles emitted from a given source, or from proximate sources, are so correlated in space-time, or momentum-energy, as to imply enhanced probability of emission of the particles with small relative momentum. A direct manifestation of this Bosonic stimulation,$^{1,3}$ wherein a Bose particle is preferentially scattered (stimulated scattering) into the final single particle state which has a higher pre-existing Bosonic population. This derives straightforwardly from the usual Bose factor ($n$+1) in $a^+|n\rangle = \sqrt{n+1} |n+1\rangle$. It is now easy to see that, from the point of view of the relative population statistics, this phenomenon of Bosonic stimulation maps on to that associated with the Polya urn scheme.$^4$ Recalling briefly, here one begins with one black and one white ball in the urn. The rule of the game then is to draw a ball at random from the urn, note its colour, and then replace the ball so drawn back into the urn, but with one additional ball of the same colour. The relative populations of the black and the white balls fluctuate initially, but eventually iterate away to a limiting value for the given run deterministically $-$ i.e., with the fluctuations about the limiting value staying within arbitrarily small tolerance chosen as $n \rightarrow \infty$, where $n$ numbers the draws. The limiting value, however, varies randomly from one run to another $-$ hence the irreproducibility of the limiting non-degenerate value. It is the {\em vagaries} of the initial few draws that determine the limiting value for a given run. Now, it is clear from the above that the rule to {\em replace the ball so drawn back into the urn, but with one additional ball of the same colour} simulates the ($n$+1) factor of the Bosonic stimulation. Its relation to a fragmentation of a BEC is now clear.

More precisely, consider a Bose-Einstein condensate as a source from which the Bosons could be coupled out into two equivalent but well separated traps. Bosonic stimulation then determines the occupation numbers of the two traps, as given by
\begin{equation}
P_{t+1} (n_1, n_2) = P_t(n_1-1, n_2)\left(\frac{n_1}{n_1+n_2+1}\right) + P_t 
(n_1, n_2-1)\left(\frac{n_2}{n_1+n_2+1}\right) ,
\end{equation}
where $P_t(n_1,n_2)$ is the probability of occupation numbers ($n_1, n_2$) for the two traps (1,2) at time {\em t}. The time here is discretized so that one particle is out-coupled in the unit time step.

This is the probabilistic description of the Polya urn scheme. Clearly, as $t\rightarrow \infty$, one settles down to relative populations  distributed uniformly in the interval zero to one. The result readily generalizes to more than two traps $-$ multifragmentation.

Fragmented B-E condensate is normally understood in terms of repulsive inter-particle interaction. As shown by Nozieres$^3$, for a repulsive inter-particle interaction the Bosonic exchange makes a fragmentation of the condensate energetically costly in equilibrium.  What we are discussing here, however, is the purely quantum statistical effect of Bosonic stimulation on the process of fragmentation $-$ a partitioning into the two traps which do not interact dynamically. The non-degenerate irreproducible relative populations are a type of partitioning noise $-$ but from one run to another.

Yet another experimental possibility is to consider a degenerate, two-mode laser which is pulsed. Here again, the photon released from the source (a stimulated emission from inverted atomic population) can populate the two degenerate cavity modes, from which the photons are coupled out eventually. Exactly the same considerations apply, with the only difference that in a pulsed laser the pulse-to-pulse fluctuations give directly the statistics of the limiting non-degenerate values of the relative populations. That should make it experimentally easier to realize.

Finally, we note that our work addresses an aspect of the Bose-Einstein correlation that enters many other fields $-$ including astronomy, the celebrated Hanbury-Brown and Twiss$^{2,5}$ (HBT) effect. The dilute gaseous BEC offers new possibilities here.

The author would like to thank Hema Ramachandran and Andal Narayanan for discussion on possible experimental demonstration of the above effect.  It is a pleasure to thank Robert M. May for a discussion on the Polya urn scheme.


\begin{thebibliography}{10}
\bibitem{}See D.W. Snoke and Gordon Baym, {\em in} ``Bose-Einstein Condensation", eds. A. Griffin, D.W. Snoke and S. Stringari (Cambridge University Press, Cambridge, 1995), p.1.
\bibitem{}G.J. Troup, ``Progress in Quantum Electronics", Vol. 2, eds. J.H. Sanders and S. Stenholm (Pergamon, Oxford, 1972).
\bibitem{}P. Nozieres, {\em in} Ref.1, p. 15.
\bibitem{}Joel E. Cohen, {\em BioScience}  {\bf 26}, 391 (1976), See also, Robert M. May, {\em Nature} {\bf 262}, 646 (1976).
\bibitem{}R. Hanbury-Brown, ``Photons, Galaxies and Stars", (Indian Acad. Sci., Bangalore, 1985).
\end{thebibliography}
\end{document}